\documentclass[twocolumn,prl,superscriptaddress]{revtex4-2}
\usepackage{amsmath,amssymb,mathrsfs}
\usepackage{natbib}
\usepackage{subfigure}
\usepackage{tabularx}
\usepackage{epsfig}
\usepackage{longtable}
\usepackage{amsfonts}
\usepackage{rotating}
\usepackage{bbold}
\usepackage{hhline}
\usepackage{braket}
\usepackage{txfonts, comment}
\usepackage{ulem}

\usepackage[unicode=true,bookmarks=true,bookmarksnumbered=false,bookmarksopen=false,breaklinks=false,pdfborder={0 0 1},backref=false,colorlinks=true]{hyperref}

\hypersetup{linkcolor=magenta,urlcolor=blue,citecolor=blue,pdfstartview={FitH},hyperfootnotes=false,unicode=true}

\def\be{\begin{equation}}
\def\ee{\end{equation}}
\def\bea{\begin{eqnarray}}
\def\eea{\end{eqnarray}}

\def\tbf{\textbf}

\begin{document}
\title{Quantum transport of strongly interacting fermions in one dimension at far-out-of-equilibrium }

\author{Jie Zou}
\affiliation{State Key Laboratory of Surface Physics, Institute of Nanoelectronics and Quantum Computing,
and Department of Physics, Fudan University, Shanghai 200433, China
}

\author{Xiaopeng Li}
\email{xiaopeng\_li@fudan.edu.cn}
\affiliation{State Key Laboratory of Surface Physics, Institute of Nanoelectronics and Quantum Computing,
and Department of Physics, Fudan University, Shanghai 200433, China
}
\affiliation{Shanghai Qi Zhi Institute, AI Tower, Xuhui District, Shanghai 200232, China}

\begin{abstract}
In the study of quantum transport, much has been known for dynamics near thermal equilibrium. 
However, quantum transport far away from equilibrium is much less well understood---the linear response approximation does not hold for physics far-out-of-equilibrium in general. 
In this work,
motivated by recent cold atom experiments on probing quantum many-body dynamics of a one-dimensional XXZ spin chain, we study the strong interaction limit of the one-dimensional spinless fermion model, which is dual to the XXZ spin chain.
We develop a highly efficient computation algorithm for simulating the non-equilibrium dynamics of this system exactly, and examine the non-equilibrium dynamics starting from a density modulation quantum state. 
We find ballistic transport in this strongly correlated setting, 
and show a plane-wave description emerges at long-time evolution. 
We also observe sharp distinction between transport velocities in short and long times as induced by interaction effects, and provide a quantitative interpretation for the long-time transport velocity.
\end{abstract}

\date{\today}
\maketitle

{\it Introduction.---}
Transport in low dimensional quantum systems~\cite{bertini2021_RMP} has been attracting continuous research interests from both theoretical and experimental perspectives, due to both the existence of solvable models~\cite{takahashi2005_solvableModel} and intrinsic strong correlation effects~\cite{bertini2021_RMP, bulchandani2021_superdiffusion}.
Although wavefunctions are exactly solvable via Bethe ansatz (BA)~\cite{giamarchi2003} in certain one-dimensional models, 
it is challenging to compute dynamical observables such as transport properties with analytic methods. 
Novel phenomenon unveiled by recent studies still demand advanced numerical techniques to explain~\cite{hild2014_PRL, Ronzheimer_PRL2013, jepsen2020_nature}.

The spin-$1/2$ XXZ chain in one dimension is a representative spin model, with the Hamiltonian
\be
H = J\sum_{i=1}^{L}(S_i^x S_{i+1}^x+ S_i^y S_{i+1}^y + \Delta S_i^z S_{i+1}^z ),
\label{eq:XXZ}
\ee
where $L$ is the system size, $J$ denotes the hopping strength and $\Delta$ labels the spin anisotropy.
Despite its simplicity, this model exhibits several interesting transport properties worth theoretical investigation.
New concepts in nonequilibrium dynamics in closed quantum systems~\cite{eisert2015_np_nonequilibrium, D'Alessio2016},
including generalized Gibbs ensemble~\cite{Rigol_GGE2007}, can also be studied in this model with a quantum quench setup~\cite{rigol2008thermalization}. 
Another reason is from the perspective of phenomenological transport laws.
The existence of anomalous transport at finite temperatures, which does not obey Fourier’s law~\cite{narasimhan1999fourier}, has been proved via linear-response theory~\cite{zotos1997_PRB_LowerBound} in this model.

Previous studies on one-dimensional quantum spin transport in the linear response regime have provided important insights about the correlated quantum dynamics.
It has been proved that finite-temperature transport is ballistic for $|\Delta| < 1 $ (gapless phase) at zero magnetization ($m_z = \sum_i \braket{s_i^z} / L= 0$)~\cite{prosen2011_PRL_gaplessBound} or for generic $\Delta$ at finite magnetization~\cite{castella1995_PRL_drude,zotos1997_PRB_LowerBound}.
For the former, recent studies suggest~\cite{ilievski2017_PRL_GHDgapped, Karrasch2014_PRB_tDMRG} a diffusive transport in gapped phase ($|\Delta| > 1 $) and predict~\cite{superdiffusive_PRL_2020} a superdiffusive transport in the Heisenberg limit ($|\Delta| \rightarrow 1 $) when $m_z = 0$.
For the latter, Mazur inequality~\cite{MAZUR} gives a finite lower bound~\cite{castella1995_PRL_drude,zotos1997_PRB_LowerBound} for the spin Drude weight $D_W^{(s)}$~\cite{scalapino1993_PRB_drudeWeight}, a criterion of ballistic transport. 
The theoretical study at high temperature~\cite{zotos1997_PRB_LowerBound} finds
\be
D_W^{(s)} \ge \frac{J^2 \Delta^2 m_z^2}{4T} \frac{1-m_z^2}{1+\Delta^2 (2+2m_z^2)} > 0.
\label{eq:Mazur}
\ee

In contrast to the linear-response regime, dynamics far-out-of-equilibrium is less well studied. 
The results obtained via linear-response theory may not hold at far-out-of-equilibrium.
Numerical methods, such as the commonly used density matrix renormalization group (DMRG) algorithm~\cite{schollwock2011_DMRG}, tend to be limited in simulating long time quantum dynamics due to the rapid entanglement growth.
Finding solvable models or efficient algorithms is thus important for investigating far-out-of-equilibrium dynamics. 

The study of far-out-of-equilibrium quantum transport becomes more desirable with recent developments in cold atom experiments. 
Experiments based on such clean and controllable closed quantum systems~\cite{bloch2008_RMP_ultracoldAtom} are naturally suitable to study the nonequilibrium dynamics, stimulating much recent research efforts in this direction ~\cite{fukuhara2013_nature_magnon, hild2014_PRL, jepsen2020_nature}.
Notably, a recent experiment~\cite{jepsen2020_nature} demonstrates a subdiffusive transport for spin helix state in a gapped phase at zero magnetization for the anisotropic Heisenberg model, instead of a diffusive behavior as expected from linear-response theory analysis~\cite{ilievski2017_PRL_GHDgapped}.
The experiment on the isotropic case also shows a violation of the linear-response approximation when considering high-energy-density spin spiral states~\cite{hild2014_PRL}.
Modeling and understanding the far-out-of-equilibrium quantum transport of the one dimensional spin chain demands further theoretical investigation. 

% Fig 1
\begin{figure*}[htp]
\includegraphics[width=0.97\linewidth]{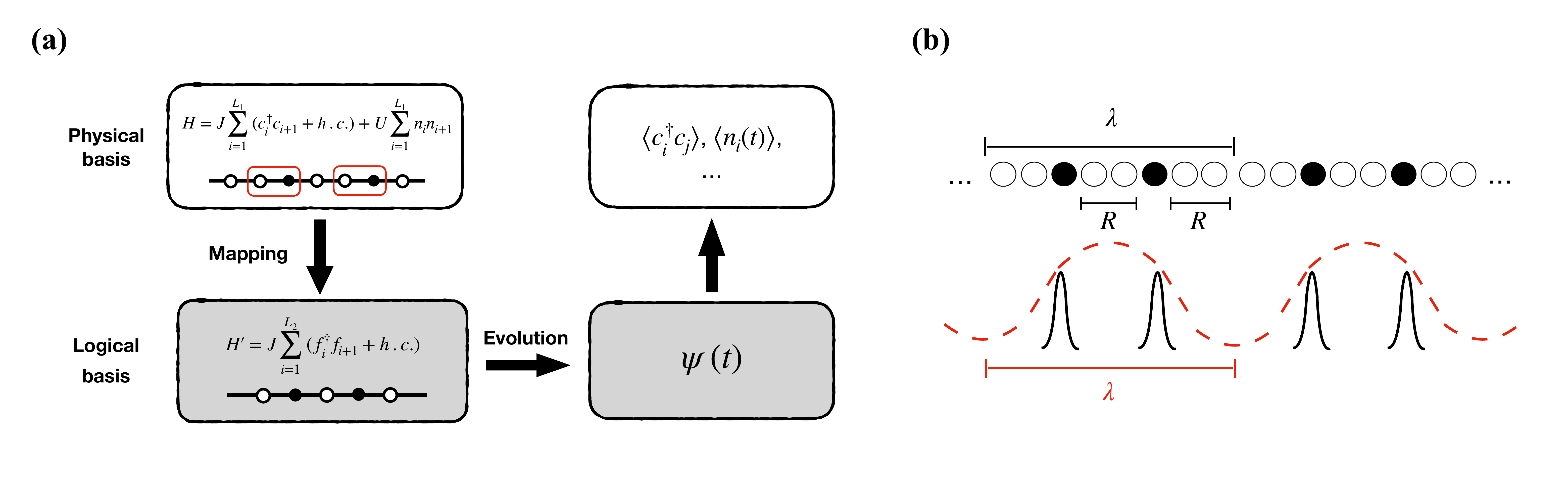}
\caption{(\tbf{a}) Workflow of the algorithm solving the strongly interacting model. The dynamics can be efficiently computed in logical basis by mapping the interacting model to a free one.
The physical observables are computed by mapping back to interacting basis~\cite{gomez1993_PRL_exactLuttinger,li2019_PRA_algorithm}, for which the key is to sample the Slater determinant wavefunction efficiently. Here we implement the recently developed fast fermion sampling algorithm~\cite{our_algorithm_2021}, in order to simulate large system sizes.
(\tbf{b}) Illustration of the density modulation state. In this state, the particles are distributed in a periodic way with a period, $\lambda$, and are separated by a distance, $R$. 
}
\label{fig:illustrate}
\end{figure*}

In this work, we focus on a one-dimensional spinless fermion model, which is equivalent to the spin-$1/2$ XXZ model. We take a strong interaction limit, in which physical observables can be efficiently calculated with our fast fermion sampling algorithm~\cite{gomez1993_PRL_exactLuttinger,li2019_PRA_algorithm, our_algorithm_2021} 
However, due to the infinitely large repulsion, we can only deal with cases away from half-filling (zero magnetization). We choose the density modulation state (DMS), a highly excited state, to check whether linear response results still hold in the far-out-of-equilibrium setting. The setup of DMS state is demonstrated in Fig.~\ref{fig:illustrate}(b).
Since the average energy of DMS corresponds to that of an infinite-temperature ensemble, the lower bound of spin Drude weight (Eq.~\eqref{eq:Mazur}) vanishes. 
In spite of this vanishing lower bound, our results show that the system still exhibits ballistic transport. Furthermore, we provide a plane-wave picture in describing the long-time dynamics of this highly excited state in the strongly interacting model. At last, we observe different transport velocities in short and long times caused by interaction effects. The transport velocity in the long-time limit is described by our plane-wave picture.

{\it Model and Methods.---}
We consider a strongly interacting fermionic model on a one-dimensional lattice, with particle number $N$ and the site index $i \in [1,L]$.
The quantum many-body dynamics is described by a Hamiltonian, 
\be
H = J\sum_{i=1}^{L_1}(c^{\dagger}_i c_{i+1}+h.c.)+U\sum_{i=1}^{L_1}\sum_{k=1}^{R}n_in_{i+k},
\label{eq:Ham}
\ee
with $i$ the lattice site index,  $c_i$ the fermionic annihilation operator,  and $n_i=c_i^\dag c_i$ the occupation number. 
Here we take the strongly interacting limit $U \rightarrow +\infty$ and $J$ is the tunneling strength. The interaction range $R$ and the filling $\rho = N/L$ are both tunable parameters in our model. In this work, we choose an open boundary condition and set the lattice constant and tunneling to be the length and energy units, respectively. In fact, this fermionic model is equivalent to a spin-1/2 XXZ model through a Jordan-Wigner transformation~\cite{giamarchi2003}. Notably, all the dynamics are suppressed in this strong interaction limit when the system is at half-filling, or equivalently at zero magnetization in the corresponding spin model.
Therefore, in this work, we carry out our theoretical study away from half-filling.

To overcome the exponential complexity of the interacting many-body system, here we adapt an efficient sampling algorithm~\cite{li2019_PRA_algorithm, our_algorithm_2021} to compute the time evolution, with an $O(M_s LN^2)$ complexity and an error $\delta$ proportional to the inverse square root of $M_s$, the number of random samples. The illustration of this method is shown in Fig.~\ref{fig:illustrate}(a). 
Since we take the infinite interaction limit, those states with two particles in a neighborhood of size $R$ are prohibited. 
Taking this constraint into account, we can map our model to an exactly solvable free fermion case~\cite{gomez1993_PRL_exactLuttinger,li2019_PRA_algorithm}.
This is realized by considering one occupied site and $R$ empty sites on its left as a a composite fermion. 
Through this mapping, the particle number remains the same as the original model and the system size is reduced down to $L_2 = L_1 - R(N-1)$.  
Hereafter, we refer to the model before (after) the mapping as the one in physical (logical) basis. Furthermore, an advanced fermion sampling method benefits us with less time cost when solving the free fermion problem. Contrary to the commonly used density matrix renormalization group (DMRG)~\cite{schollwock2011_DMRG} method, which only allows an investigation on systems up to hundreds of sites in short and intermediate times, our algorithm can solve the dynamics exactly regardless of the evolution time, for a large system with up to thousands of sites.

Taking the interacting fermion model, we consider the time evolution of a density modulation state, which are analogous to spin helix states investigated in recent experiments~\cite{hild2014_PRL,jepsen2020_nature} through Jordan Wigner transformation. 
It is a highly excited state since the expectation value of its energy is equal to zero, corresponding to the infinite temperature ensemble average. 
The preparation of this state is shown in Fig.~\ref{fig:illustrate}(b). Fermions are arranged in a periodic structure with $\lambda$ the wavelength, and thus the particle number in a period is determined for a given filling $\rho$. Within a period, these particles are aligned in a most compact pattern, that is, the distance of two adjacent particles is set to be the interaction range $R$. 
At last, a suitable system size is also necessary to minimize the finite size effect and the computation cost, which we set $L=4\lambda$ in this work.

% Fig 2
\begin{figure*}[htp]
\includegraphics[width=.8\linewidth]{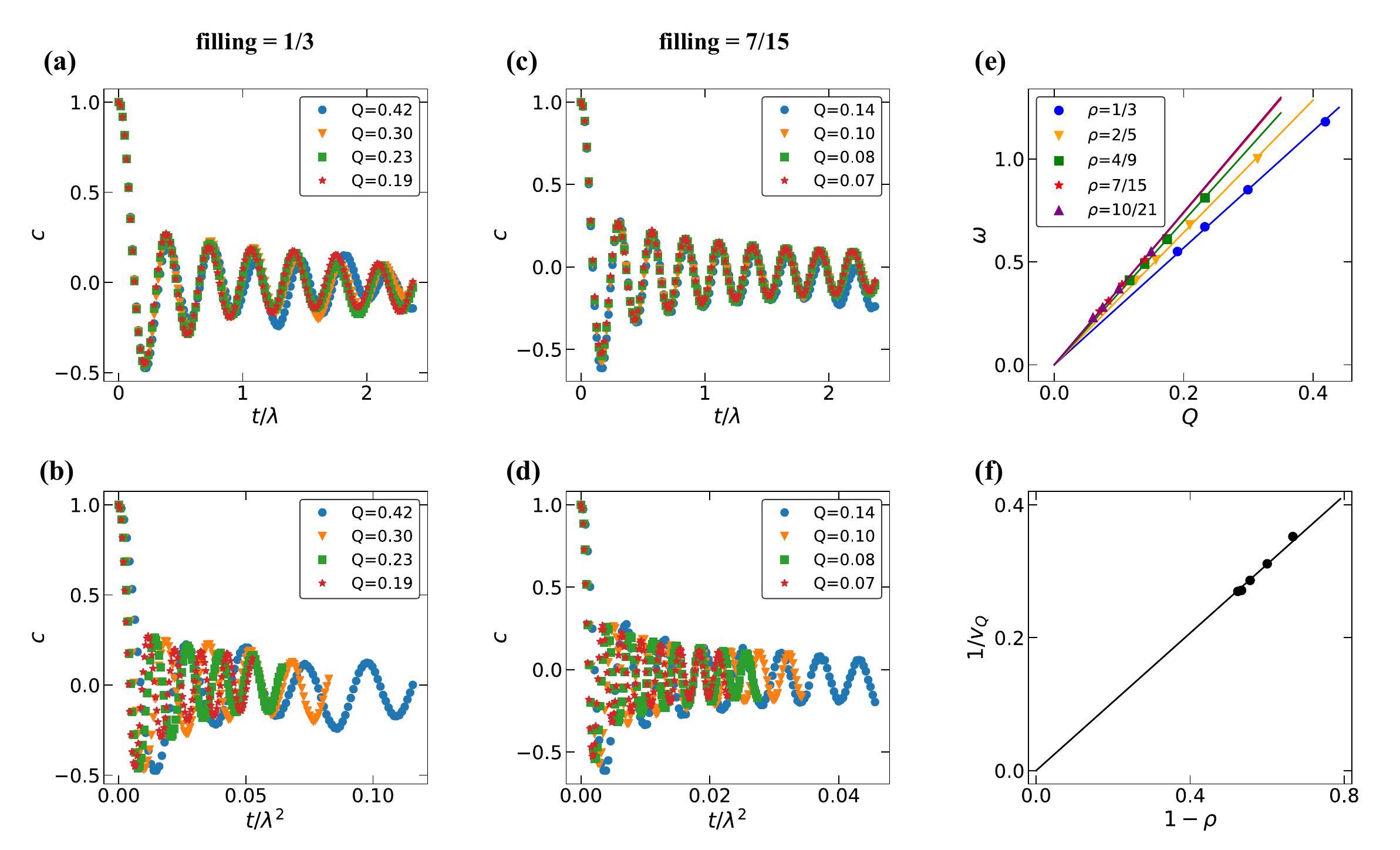}
\caption{Relaxation of the density modulation state. The interaction range here is fixed to $R=1$. (\tbf{a}, \tbf{b}), the normalized contrast, $c(t)$ in Eq.~\eqref{eq:normcontrast}, with different wavevectors ($Q$), at filling  $\rho=1/3$. 
The evolution time $t$ is rescaled by the wavelength $\lambda$ and its square $\lambda^2$ in ({\bf a}) and ({\bf b}), respectively. 
(\tbf{c}, \tbf{d}), 
the time evolution of the normalized contrast at a different filling, $\rho=7/15$. 
(\tbf{e}) Long-time oscillation frequency ($\omega$) of the contrast as a function of the wavevector ($Q$). 
The numerical data points are well fitted by a linear function $\omega = v_Q Q$, with $v_Q$ the dispersion velocity.
The slopes $v_Q$ are 2.84, 3.21, 3.49, 3.69, and 3.71 for filling factors $\rho = 1/3$, $2/5$, $4/9$, $9/15$, and $10/21$.  
(\tbf{f}), the dependence of $v_Q$ on the filling. 
A nontrivial linear relation between the inverse of dispersion velocity and the filling factor is uncovered. We find 
$1/v_Q \approx 0.52 \times (1-\rho)$ by an empirical fitting.
}
\label{fig:relaxation}
\end{figure*}

{\it Relaxation.---}
The relaxation dynamics is characterized by examination of the density modulation amplitude. 
Due to the periodicity of density modulation states, the contrast of density modulation is defined through a Fourier transform  
\be
C(t)=\frac{2}{L}\sum_{i=1}^L \braket{n_i(t)} \cos(Qi),
\label{eq:contrast}
\ee
where $Q = 2\pi / \lambda$ is the wavevector. For comparison, we focus on the normalized contrast 
\be 
c(t) = C(t)/C(0).
\label{eq:normcontrast} 
\ee 
This physical observable is also commonly used to characterize a spin helix state in recent experiments~\cite{jepsen2020_nature}. 

% Fig 3
\begin{figure}[tph]
\includegraphics[width=\linewidth]{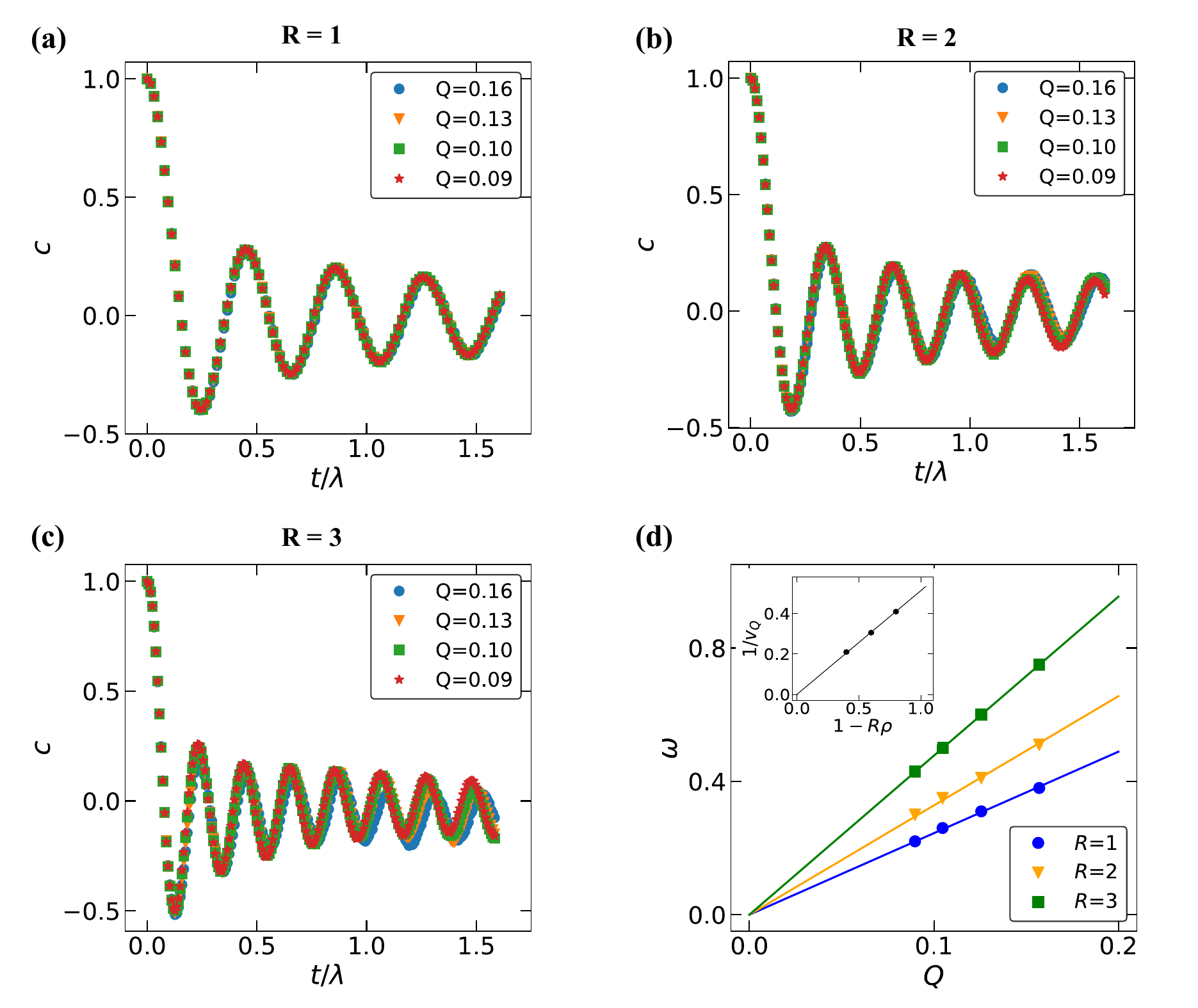}
\caption{
The dependence of density modulation relaxation dynamics on the interaction range. 
(\tbf{a}, \tbf{b}, \tbf{c}), the normalized contrast $c(t)$ (Eq.~\eqref{eq:normcontrast}) with interaction ranges $R=1$, $R=2$, and $R=3$. We investigate the relaxation dynamics for a broad range of wavevector ($Q$) from $0.09$ to $0.16$. 
%with different wavevector $Q$s. The interaction range varies from 1 to 3 and the filling $\rho$ is fixed to $1/5$. 
The evolution time $t$ is rescaled by the wavelength $\lambda$. (\tbf{d}),  long-time oscillation frequency of the contrast. 
The numerical data points are all captured by the linear relation $\omega = v_Q Q$. 
The dispersion velocities $v_Q$, are 2.44, 3.28, and 4.77, for the interaction range $R=1$, $2$, and $3$, respectively. Its inset shows the linear relation between the inverse of the dispersion velocity $v_Q$ and the interaction range $R$. The filling $\rho$ is fixed to $1/5$ here. 
} 
\label{fig:range}
\end{figure}

In the linear response regime, whether a system is half-filled plays a key role in its transport property~\cite{zotos1997_PRB_LowerBound, prosen2011_PRL_gaplessBound}.
One of the main purposes of our work is to investigate the strong interaction effects on the quantum transport. 
We set the filling $\rho = \frac{m}{2m+1}\,(m=1,2,...)$ so that we can gradually probe the half-filling limit. The results are shown in Fig.~\ref{fig:relaxation}. 
We observe a perfect data collapse in the density wave contrast ($C(t)$ in Eq.~\eqref{eq:contrast}), with the evolution time ($t$) rescaled by the wavelength $\lambda$. 
This data collapse holds even for a filling factor quite close to $1/2$ ($\rho = 7/15$).
This data-collapse behavior is absent if the time $t$ is rescaled by the square of the wavelength $\lambda^2$ instead. These numerical results imply universal ballistic transport in this model, despite the strong interaction induced nonlinearity. 

\begin{figure}[!t]
\includegraphics[width=\linewidth]{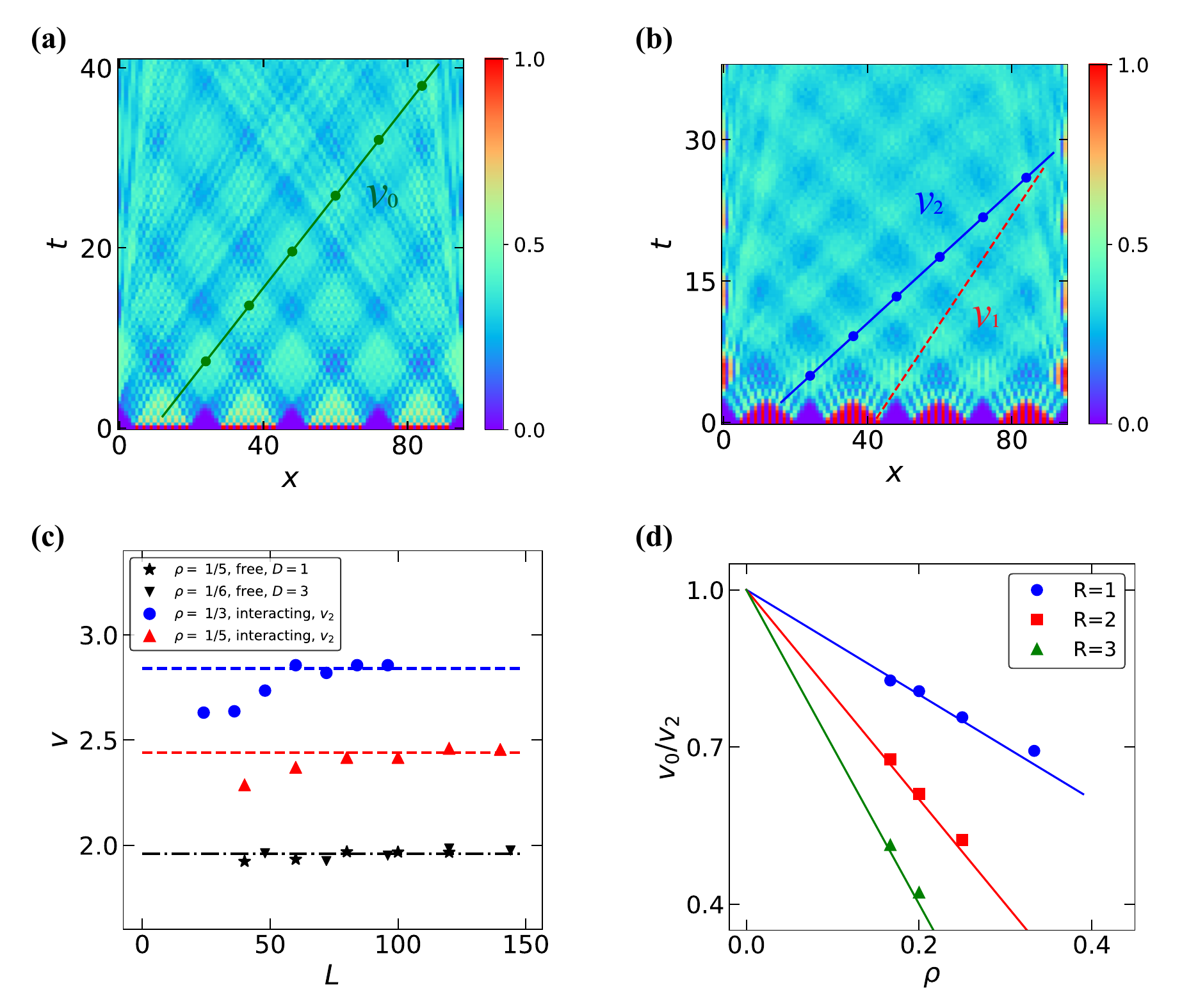}
\caption{Sound velocity renormalization. ({\bf a}) and ({\bf b}) show the time evolution of the density profile $\braket{n_i(t)}$ for free and interacting fermions, respectively.
The sound velocities are determined according to the propagation of the wave front as marked by the green line in ({\bf a}), and the blue and red lines in ({\bf b}). 
The free fermion dynamics is characterized by one velocity scale, $v_0$, whereas the two different velocities, $v_1$ and $v_2$, appear in the early and late stages of the interacting dynamics. 
In ({\bf b}), we choose a system size $L=96$, particle number $N=32$, wavelength $\lambda = 24$. The initial state is  a density modulation state as shown in Fig.~\ref{fig:illustrate}. We use the same setup for non-interacting fermions in ({\bf a}) for a fair comparison.  
(\tbf{c}), convergence of the propagating velocities with increasing the system size. 
The velocities for the non-interacting model (black triangles and stars), with different fillings $\rho$ and particle separations within a period $D$, converge to $v=1.96$ (black dotted line). 
The velocity $v_2$ for interacting fermions (blue dots and red triangles) converges to the dispersion velocity $v_Q$ (dashed lines), as extracted from long time oscillation of the dynamics.
In ({\bf a}), ({\bf b}), and ({\bf c}), we choose $R=1$.
(\tbf{d}) The ratio between $v_0$ and $v_2$  as a function of the filling $\rho$.  The numerical results are well fitted by a linear function. 
}
\label{fig:velocity}
\end{figure}

Since the quantum transport in this model is ballistic, we make an analogue with plane wave propagating in free space with a certain velocity. 
Surprisingly, the ballistic transport that emerges from the  far-out-of-equilibrium DMS state in our strongly interacting model is captured by a physical picture of propagating plane-wave in free-space. 
For a massless plane wave mode spreading in free space with wavevector $k$ and energy $E=\hbar \omega$, the dispersion relation is $\omega = v k$, where $v$ is a dispersion velocity that determines the ballistic transport velocity of the plane wave in free space.  
In our model, we find the linear dispersion emerges $\omega_Q = v_Q Q$ in the long-time oscillation dynamics of the DMS state, for a broad range of wavevector, $Q$ (Fig.~\ref{fig:relaxation}). 
Besides, the dispersion velocity extracted from the long-time oscillation quantitatively agrees with the spreading velocity (see Fig.~\ref{fig:velocity}(c)). 
These numerical results imply the quantum dynamics of the strongly interacting fermions in one dimension starting from the far-out-of-equilibrium DMS state is well captured by a physical picture of propagating dynamics of a massless plane wave mode. 
The strong interaction effects on the dynamics are reflected by the nontrivial dependence of the dispersion velocity $v_Q$ on the filling $\rho$~(Fig.~\ref{fig:relaxation}(f)), to be further elaborated in the later analysis of velocity renormalization.

To confirm the generality of our finding, we further investigate the relaxation dynamics at different interaction ranges ($R$).
The results are shown in Fig.~\ref{fig:range}. We observe that for $R=1$, $2$, and $3$, the relaxation dynamics starting from DMS states with different wavelengths ($\lambda$) all collapse on a curve when time $t$ is rescaled by $\lambda$.   
This confirms the ballistic transport for the different choices of interaction ranges. We find that the dispersion velocity extracted from late time oscillation dynamics becomes larger as we increase the interaction range $R$ (Fig.~\ref{fig:range}(d)). This can be attributed to that longer interaction range in our model tends to accelerate the dynamical transport.

% Fig 4

{\it Velocity Renormalization.---}
Velocity is a characteristic property of ballistic transport, as well as an observable which can be measured directly in experiments. 
In our numerical simulation here, we determine the sound velocity by extracting the slope of the stripe pattern in the density profile $\braket{n_i(t)}$.
A typical time evolution of $\braket{n_i(t)}$ is shown in Fig.~\ref{fig:velocity}(a) ((b)), considering the density modulation state in the free (strongly interacting) model. The free fermion velocity $v_0$ and the short-time (long-time) velocity $v_1$($v_2$) in the interacting case is also determined by fitting the stripe pattern in Fig.~\ref{fig:velocity}(a) and (b).
The strongly interacting model shows different behaviors at early and late time evolution. 
This two-stage dynamical feature as originated from strong interaction effects makes the system distinctive from noninteracting fermions.  
Moreover, even at late time where the ballistic transport of the strongly interacting model looks similar to the non-interacting case, the propagating velocity is actually strongly renormalized by the interaction. 

To gain more insight into the renormalization process, we quantitatively study the velocities mentioned above.
As system size grows, the free fermion velocity $v_0$ and the long-time velocity $v_2$ in the interacting case converge, which is demonstrated in Fig.~\ref{fig:velocity}(c).
For comparison, we also choose DMS states as initial states in the non-interacting case. Again, the particles are aligned in a periodic way (see Fig.~\ref{fig:illustrate}(b)), with $D$ the distance between particles within a period, in analog with $R$ in the interacting case.
It is shown that the converged value of $v_0$ is independent with $D$ and filling $\rho$, as expected for free fermions. 
Hereafter, the velocities $v_0$ and $v_2$ are defined according to their thermodynamic limits.
We observe that the velocity $v_2$ matches the dispersion velocity $v_Q$ extracted from the long-time oscillations. 
The linear fit in Fig.~\ref{fig:velocity}(d) suggests the relation $v_0/v_2 = 1-R\rho$. This is verified in our numerical calculation for a broad range of $R$ and $\rho$. 

The renormalization of the velocity in the interacting dynamics as compared to the non-interacting case can be understood from the mapping between the logical and physical basis. In this mapping, a length scale ($l$) in the physical basis (interacting)  shrinks to $l-l R\rho$ in the logical basis (non-interacting) on average. Such shrinking has been used in the previous study on the ground state of the 1D strongly interacting fermions for the Luttinger parameter~\cite{gomez1993_PRL_exactLuttinger}. 
In our study of the dynamics, since the long-time behavior of the strongly interacting model is captured by the plane-wave picture, 
it is reasonable to expect the velocity renormalizes according to the length scale, which then implies 
\be
\frac{v_0}{v_2}  \simeq 1-R\rho. 
\label{eq:velo}
\ee
This is exactly the same as what we observed in our numerical simulations.

{\it Conclusion.---}
In this work, we study the quantum transport properties of strongly interacting spinless fermions in one dimension. This model is simulated efficiently using a fast fermion sampling algorithm at the strong interaction limit. Despite the strong interaction, we observe robust ballistic transport for different fillings and interaction ranges. We propose a plane-wave description for the ballistic transport of the strongly interacting fermions, which captures the long-time dynamics.  The predominant effect of the strong interaction on the transport is to introduce velocity renormalization. 
Our results indicate the subdiffusive transport observed for the weakly interacting spin chain in the experiment~\cite{jepsen2020_nature} would have a crossover to  ballistic behavior when approaching the  strongly interacting regime. 
This is worth further experimental investigation.

{\it Acknowledgement.---}
We acknowledge support by National Program on Key Basic Research Project of China (Grant No. 2021YFA1400900), National Natural Science Foundation of China (Grants No. 11934002), Shanghai Municipal Science and Technology Major Project (Grant No. 2019SHZDZX01), Shanghai Science Foundation (Grants No.21QA1400500, 19ZR1471500).

\bibliography{references}

\end{document}